\documentstyle[aps]{revtex}
\begin{document}
\tightenlines
\title{Gauge - invariant
fluctuations of the metric in stochastic inflation}
\author{Mauricio Bellini\footnote{E-mail address: mbellini@mdp.edu.ar}}
\address{Departamento de F\'{\i}sica, Facultad de Ciencias Exactas  y
Naturales \\ Universidad Nacional de Mar del Plata, \\
Funes 3350, (7600) Mar del Plata, Buenos Aires, Argentina.}
\maketitle
\begin{abstract}
I derive the stochastic equation for the perturbations of the metric
for a gauge-invariant energy - momentum-tensor (EMT) in stochastic
inflation. A quantization for the field that describes the
gauge-invariant perturbations for the metric is developed.
In a power - law expansion for the universe the amplitude for these
perturbations on a background metric
could be very important in the infrared sector.
\end{abstract}
\twocolumn
During inflation vacuum fluctuations on scales less than the
Hubble radius are magnified into classical perturbations in the scalar
fields on scales larger than the Hubble radius. These classical
perturbations in the scalar fields can then change the
number of e-folds of expansion and so lead to classical
curvature and density perturbations after inflation. These
density perturbations are thought to be responsible for the formation
of galaxies and the large scale structure of the observable universe
as well as, in combination with the gravitational waves produced
during inflation, for the anisotropies in the cosmic microwave
background.

In this report I consider the gauge-invariant fluctuations of the
metric on a globally flat Friedmann - Robertson - Walker
(FRW) metric in the early inflationary universe.
These metric fluctuations are here considered in the framework
of the linear perturbative corrections.
A non - linear
perturbative calculation for this issue was developed
in\cite{1,2}.
The scalar metric perturbations of the metric are associated
with density perturbations. These are the spin - zero projections of
the graviton, which only exist in non - vacuum cosmologies.
The issue of gauge invariance becomes critical when we attempt
to analyze how the scalar metric perturbations produced in the very
early universe influence of a globally flat isotropic and
homogeneous universe. This allows to formulate the problem of the
amplitude for the scalar metric perturbations on the evolution
of the background FRW universe in a coordinate - independent
manner at every moment in time.
Since the results do not depend on the gauge, the perturbed
globally flat
isotropic and homogeneous universe is well described by\cite{3}
\begin{equation}\label{m}
ds^2 = (1+2\psi) \  dt^2 - a^2(t) (1-2\chi) \  dx^2,
\end{equation}
where $a$ is the scale factor of the universe and $\psi$ and $\chi$
the perturbations of the metric. I will consider the particular
case where the tensor $T_{ij}$ is diagonal, i.e., for $\chi = \psi$\cite{4a}.
As in a previous work\cite{4} I consider a semiclassical expansion for the
scalar field $\varphi(\vec x,t) = \phi_c(t) + \phi(\vec x,t)$,
with expectation values
$<E|\varphi|E> = \phi_c(t)$ and $<E|\phi|E>=0$. Here,
$|E>$ is a unknown state of the universe.
Due to $<E|\chi|E> =0$, the expectation
value of the metric (\ref{m}) gives the background
metric that describes a flat FRW spacetime.
Linearizing the Einstein
equations in terms of $\phi$ and $\chi$, one obtains the system of
differential equations for $\phi$ and $\chi$
\begin{eqnarray}
 \ddot\chi &+& \left(\frac{\dot a}{a}
- 2 \frac{\ddot\phi_c}{\dot\phi_c} \right)
\dot \chi - \frac{1}{a^2} \nabla^2 \chi   \nonumber \\
&+&\left[
\frac{\ddot a}{a} - \left(\frac{\dot a}{a}\right)^2 - \frac{\dot a}{a}
\frac{\ddot\phi_c}{\dot\phi_c}\right] \chi =0, \label{1}\\
\frac{1}{a}& \frac{d}{dt}& \left( a \chi \right)_{,\beta} =
\frac{4\pi}{M^2_p} \left(\dot\phi_c \phi\right)_{,\beta} , \\
\ddot\phi& +& 3 \frac{\dot a}{a} \dot\phi -
\frac{1}{a^2} \nabla^2 \phi + V''(\phi_c) \phi \nonumber \\
& + & 2 V'(\phi_c) \chi- 4 \dot\phi_c \dot\chi =0.
\end{eqnarray}
Here, the dynamics of $\phi_c$ being described by the equations
\begin{eqnarray}
&& \ddot\phi_c + 3 \frac{\dot a}{a} \dot\phi_c + V'(\phi_c) = 0, \\
&& \dot \phi_c = - \frac{M^2_p}{4 \pi} H'_c(\phi_c),
\end{eqnarray}
where the prime denotes the derivative with respect to $\phi_c$ and
$H_c(\phi_c) \equiv {\dot a\over a}$. The equation (\ref{1})
for $\chi$
can be simplified with the map $h =
e^{1/2 \int \left({\dot a \over a}-
{2 \ddot\phi_c \over \dot\phi_c} \right) dt} \  \chi$
\begin{eqnarray}
\ddot h &-& \frac{1}{a^2} \nabla^2 h +  \left[ \frac{1}{4}
\left(\frac{\dot a}{a} - 2 \frac{\ddot \phi_c}{\dot\phi_c} \right)^2
\right.\nonumber \\
& -&\left. \frac{1}{2} \left( \frac{\ddot{a} a - \dot{a}^2}{a^2}
- \frac{2 \frac{d}{dt}\left(\ddot\phi_c \dot\phi_c\right)
- 4 \dot\phi^2_c}{\dot\phi^2_c} \right)\right.\nonumber\\
& +& \left( \frac{\ddot a}{a} - \left( \frac{\dot a}{a}\right)^2
- \left.
\frac{\dot a}{a} \frac{\ddot\phi_c}{\dot\phi_c} \right) \right] h
= 0\label{2}.
\end{eqnarray}
The eq.(\ref{2}) is a Klein - Gordon equation for the
redefined fluctuations of the metric $h(\vec x,t)$ in a curved
spacetime defined with a flat FRW metric for the background.
This field can be written as a Fourier expansion in terms of the
modes $h_{\vec k} =e^{i \vec k. \vec x} \xi_k(t)$
\begin{equation}
h(\vec x,t) = \frac{1}{(2\pi)^{3/2}} \int
d^3 k \left[ a_k h_k + a^{\dagger}_k h^*_k\right],
\end{equation}
where $a_k$ and $a^{\dagger}_k$ are the annihilation and creation
operators with commutation relations $[a_{\vec k},a^{\dagger}_{\vec k'}] =
\delta^{(3)}(\vec k - \vec k')$, and the asterisk denotes the complex
conjugate.
The matter field perturbations, written as a Fourier expansion, is
\begin{equation}
\phi(\vec x,t) = \frac{1}{(2\pi)^{3/2}} \int
d^3 k \left[ a_k \phi_k + a^{\dagger}_k \phi^*_k\right],
\end{equation}
where $\phi_{\vec k} =e^{i \vec k. \vec x} u_k(t)$ .
Due to the fact $\chi = \psi$, the metric and matter perturbations
are anticorrelated outside the horizon: $\xi_k
=- \phi_c(t) \  e^{-1/2 \int \left({\dot a \over a}-
{2 \ddot\phi_c \over \dot\phi_c} \right) dt} \  u_k$\cite{4a}.
The equation for the time dependent modes $\xi_k(t)$
is
\begin{equation}\label{tm}
\ddot\xi_k + \omega^2_k(t) \xi_k =0,
\end{equation}
where $\omega_k(t)$ is
the time dependent frequency for each mode with wavenumber $k$:
$\omega^2_k(t) = [k^2/a^2 - k^2_o/a^2]$. Here, $k_o(t)$ is
the time dependent wavenumber that separates the infrared (IR)
and the ultraviolet (UV) sectors. On super Hubble scales,
$k^2/a^2 \ll k^2_o/a^2$ and the equation (\ref{tm})
for the time dependent
frequencies become $k$-independent:
$\omega^2_k(t) \simeq - k^2_o(t)/a^2$.
The commutation relation for $h$ and $\dot h$ is
$[h(\vec x,t),\dot h(\vec x',t)]= {\rm i}
\delta^{(3)} (\vec x - \vec x')$ for
$\xi_k \dot\xi^*_k - \dot\xi_k \xi^*_k ={\rm i}$.
When the modes become real one obtains
$\xi_k \dot\xi^*_k - \dot\xi_k \xi^*_k =0$ and the
field $h$ is classical\cite{5}.

-{\em Stochastic Approach:} Now I consider the
field $h(\vec x,t)$ on the IR sector. For $k \ll k_o(t)$
the coarse - grained fields $h_{cg}(\vec x,t)$
and $\phi_{cg}(\vec x,t)$ can be written
as
\begin{eqnarray}
h_{cg}(\vec x,t)& = &
\frac{1}{(2\pi)^{3/2}} \int d^3 k \  \theta(\epsilon k_o-k)
[a_k h_k + a^{\dagger}_k h^*_k],\label{s} \\
\phi_{cg}(\vec x,t)& =& \frac{1}{(2\pi)^{3/2}}
\int d^3 k \  \theta(\epsilon k_o-k)
[a_k \phi_k + a^{\dagger}_k \phi^*_k],
\end{eqnarray}
where $\epsilon \ll 1$ is an dimensionless constant. Replacing
the eq. (\ref{s}) in (\ref{2}), one obtains the following
stochastic equation for $h_{cg}$
\begin{equation}\label{st}
\ddot h_{cg} -\frac{k^2_o}{a^2} h_{cg} = \epsilon \left[
\frac{d}{dt} \left(\dot k_o \eta \right) + 2 \dot k_o \kappa \right],
\end{equation}
where the noises $\eta$ and $\kappa$ are
\begin{eqnarray}
\eta(\vec x,t) &=& \frac{1}{(2\pi)^{3/2}} \int d^3 k \delta(\epsilon k_o - k)
[a_k h_k + a^{\dagger}_k h^*_k ], \\
\kappa(\vec x,t) &=& \frac{1}{(2\pi)^{3/2}}
\int d^3 k \delta(\epsilon k_o - k)
[a_k \dot h_k + a^{\dagger}_k \dot h^*_k ].
\end{eqnarray}
The fluctuations $h_{cg}$ will be classical if all the modes
$\xi_k$ (for $k \ll k_o$) are real\cite{5}. Thus, when
\begin{equation}
\left|\frac{{\rm Im} \left( \xi_k \right)}{{\rm Re} \left( \xi_k\right) }
\right| \ll 1,
\end{equation}
for $0 < k < \epsilon k_o$,
the eq. (\ref{st}) is a classical stochastic equation that describes
the gauge - invariant redefined fluctuations on the IR sector.
When $(\dot k_o)^2 <\kappa^2> \ll  (\ddot k_o)^2 <\eta^2>$, one
can neglect the noise $\kappa$ with respect to $\eta$ in
the eq. (\ref{st}). In this case one obtain the following two first
order stochastic equations for $h_{cg}$
\begin{eqnarray}
\dot h_{cg} & = & \epsilon \dot k_o \eta + u, \\
\dot u & = & \frac{k^2_o}{a^2} \  h_{cg},\label{sys}
\end{eqnarray}
where $u$ is an auxiliar field. The Fokker - Planck equation for the
system (\ref{sys}) gives the transition probability
$P(h_{cg},u,t|h^{(0)}_{cg},
u^{(0)}, t_o)$ for the universe, from the initial configuration
$(h^{(0)}_{cg},u^{(0)}, t_o)$ to the $(h_{cg},u,t)$ one
\begin{eqnarray}
\frac{\partial P}{\partial t} &=&- u \frac{\partial P}{\partial h_{cg}} -
\frac{k^2_o}{a^2} h_{cg} \frac{\partial P}{\partial u}
+ \frac{\epsilon^3 \dot k_o k^2_o}{4 \pi} \xi^2_{\epsilon k_o} \nonumber \\
& \times &
\left[ \frac{\partial^2 P}{\partial h^2_{cg}} \right].
\end{eqnarray}

-{\em Heisenberg representation for $h_{cg}$}:
The eq. (\ref{st}) can be written as
\begin{equation}\label{a7}
\ddot h_{cg}- \left[\frac{k_o(t)}{a(t)}\right]^2 \  h_{cg}+
\xi_c(\vec x, t)=0,
\end{equation}
where
$\xi_c(\vec x, t)=- \epsilon \left[{d \over dt}(\dot k_o \eta^{(c)})
+ 2 \dot k_o \kappa \right]$.
This noise becomes from the short wavelength sector due to the
cosmological evolution of both, the horizon and the scale factor of the
universe.
The effective Hamiltonian associated with eq. (\ref{a7}) is
\begin{equation}\label{a8}
H_{eff}(h_{cg},t)= \frac{1}{2} P^2_{cg}+
\frac{1}{2} \mu^2(t) \  \left(h_{cg}\right)^2
+ \xi_c h_{cg},
\end{equation}
where $P_{cg}\equiv \dot h_{cg}$ and $\mu^2(t)={k^2_o\over a^2}$.
Note that $\xi_c $
plays the role of an external classical stochastic
force in the effective Hamiltonian
(\ref{a8}). Thus, one can write the following Schr\"odinger equation
\begin{eqnarray}
{\rm i}\frac{\partial }{\partial t} \Psi(h_{cg},t)& = &
-\frac{1}{2} \frac{\partial^2}{\partial \left(h_{cg}\right)^2}
\Psi(h_{cg},t) \nonumber \\
& +& \left[\frac{1}{2} \  \mu^2(t) \left(h_{cg}\right)^2+
\xi_c \  h_{cg} \right] \Psi(h_{cg},t),\label{a9}
\end{eqnarray}
where $\Psi(h_{cg},t)$ is the wave function that
characterize the system.
Observe that generally $\mu(t)$ depends on time, and the
Hamiltonian (\ref{a8}) is non - conservative, also in the case in which one
would neglect the stochastic force. The only case where
$\mu$ does not present time dependence is in a de Sitter expansion
of the universe.
In this case $\mu$ is constant
and eq. (\ref{a8}) represents a harmonic oscillator with a stochastic
external force $\xi_c$.
In this case we have a forced linear harmonic
oscillator and the solution is a coherent state with the displacement
due to the action of the external force.
The
effective Hamiltonian (\ref{a8}) describes an open system. This is
due to the fact that degrees of freedom of the infrared sector
are constantly increasing since the $k_o$-temporal dependence.

The probability to find the universe with a given $h_{cg}$
in a given time $t$ is
\begin{equation}
P(h_{cg},t)=\Psi(h_{cg},t) \Psi^*(h_{cg},t),
\end{equation}
where the asterisk denotes the complex conjugate.

-{\em Power - law expansion:}
Now I study the particular case of a power - law expansion for
the universe. In this case the scale factor is
$a(t) \propto  \left({t\over t_o}\right)^p$, and the
Hubble parameter $H_c[\phi_c(t)]= p/t$. The temporal evolution
of the background field $\phi_c$ is $\phi_c(t) =
\phi^{(o)}_c - m \  {\rm ln}[(t/t_o) p)]$, where
$\phi^{(o)}_c \equiv \phi_c(t=t_o)$ --- for $t \ge t_o$.
Here, $m\simeq (10^{-4} - 10^{-6}) \  M_p$
is the mass of the inflaton field. The map for the redefined
fluctuations $h$ become $h(\vec x,t) = t^{(p/2+1)}
\chi(\vec x,t)$. Furthermore, one obtains
\begin{equation}
\frac{k^2_o}{a^2} = - \frac{1}{4} \left(p^2+6 p +4\right) t^{-2}.
\end{equation}
In the IR sector one obtains the wavelength are
greater than the size of the horizon
(i.e., $k^{-1} \gg k^{-1}_o(t)$) and
the term $k^2/a^2 $ can be neglected with
respect to the another one $k^2_o/a^2$ in the equation
for the temporal modes $\xi_k$. Hence,
the general solution for $\xi_k$ is
\begin{equation}
\xi_k(t) \simeq C_1 \  t^{1/2(1- {\rm i} \sqrt{|1-4 K^2|})} +
C_2 \  t^{1/2(1+ {\rm i} \sqrt{|1-4 K^2|})},
\end{equation}
where $K^2 = 1/4(p^2+ 6p +4)$.
It is well known that the fluctuations in the infrared sector
become classical. Thus, I will consider that the condition
$\xi_k \dot\xi^*_k - \dot\xi_k \xi^*_k =0$ holds in this sector.
This implies
that $C_1 = \pm C_2$. I consider the case where the universe expands
very rapidly ($p \gg 1$)

\begin{itemize}
\item{\em Case 1)} $C_1 = - C_2$: in this case the time dependent modes
are (for $C_1 = - {\rm i} |C_1|$)
\begin{equation}
\xi^{(1)}_k(t) \simeq 2 |C_1| \  \sin[\omega(t) t],
\end{equation}
where
\begin{equation}\label{ome}
\omega(t) = {\sqrt{|1- 4 K^2|}\over 2 t} \  {\rm ln}[t],
\end{equation}
is the frequency of the scalar perturbations of the metric, which depends on time.
\item {\em Case 2)} $C_1 =  C_2$: here the modes are (for $C_1 = |C_1|$)
\begin{equation}
\xi^{(2)}_k(t) \simeq  2 |C_1|t^{1/2} \  \left[ 1 +
t^{-1/2} \cos\left[\omega(t) t\right]\right],
\end{equation}
where $\omega(t)$ is given by eq. (\ref{ome}).
\end{itemize}
Note that ${\rm lim}_{t \rightarrow \infty} \omega(t) \rightarrow 0$ and
${\rm lim}_{t \rightarrow 0} \omega(t) \rightarrow \infty$ in both cases,
1) and 2).
The case 1) describes an oscillatory scalar perturbations of the metric
with constant
amplitude $|C_1|$ and an oscillation frequency that decreases with time.
This means that for very large $t$ the scalar perturbations of the metric
oscillates very slowly.
The case 2) describes a scalar metric perturbations with increasing
amplitude $2 |C_1| t^{1/2}$ but stop oscillating for
very large $t$.
In this case
the amplitude for the metric perturbations become very important
with time in the IR sector. Thus, in the IR sector
the amplitude for the metric fluctuations $\chi_{cg} = t^{-(p/2+1)} h_{cg}$
become
\begin{equation}
<\chi^2_{cg}>
\simeq  \frac{t^{-(p+2)}}{6 \pi^2} \int^{k_o(t)}_{0} dk \  k^2 \
\left[\xi^{(1,2)}_k(t)\right]^2,
\end{equation}
which becomes
\begin{equation}\label{a}
<\chi^2_{cg}> \propto \left[\xi^{(1,2)}_k(t)\right]^2 \  t^{p-5}.
\end{equation}
Here $\xi^{(1,2)}_k(t)$ denotes the time dependent modes $\xi_k(t)$
for the cases 1) and 2), respectively.
Note that $<\chi^2_{cg}>$ increases with time for $p>5$ in case 1),
and for $p>4$ in case 2).
Furthermore,
the density fluctuations for the matter energy density is\cite{6}
${\delta \rho \over \rho} = -2 \chi$, so that
\begin{equation}\label{b}
\left.\frac{<(\delta \rho)^2>^{1/2}}{\rho}\right|_{IR}
\propto <\chi^2_{cg}>^{1/2}.
\end{equation}
Since the metric and matter perturbations are anticorrelated
[$\xi^{(1,2)}_k = -\phi_c(t) \  t^{-(p/2+1)} u^{(1,2)}_k$],
one can write the density fluctuations in terms of $u^{(1,2)}_k$
[see eqs. (\ref{a}) and (\ref{b})].

To summarize,
a stochastic approach
for the field that describes the gauge - invariant
perturbations for the metric was developed.
These fluctuations describes an effective Hamiltonian $H_{eff}$ for
an harmonic oscillator with an effective time dependent parameter of
mass ${k^2_o \over a^2}$ and
an external stochastic force $\xi_c$.
Finally, in this report I demonstrated that
the metric fluctuations can be very important on the IR sector,
in a power - law expanding universe, when $p$ is
sufficiently large. Thus, in a power - law expansion with large
$p$, one obtains large amplitude for scalar perturbations of the metric.
\vskip .3cm
M. Bellini thank O. A. Sampayo and R. L. W. Abramo
for fruitful discussions.

\end{document}